\documentclass[twocolumn]{openjournal}
\usepackage{graphicx} 
\usepackage{multirow}

\usepackage[colorlinks,linkcolor=blue,citecolor=blue,urlcolor=blue ]{hyperref}
\usepackage[utf8]{inputenc}
\usepackage{float}
\usepackage{xcolor}
\usepackage{ulem}
\usepackage[T1]{fontenc}

\newcommand{\msolar} {$\rm{M_{\odot}}~$}
\newcommand{\msolarc} {$\rm{M_{\odot}}$}
\newcommand{\msolaryr} {$\rm{M_{\odot} / yr^{-1}}~$}
\newcommand{\msolaryrc} {$\rm{M_{\odot} / yr^{-1}}$}

\newcommand{\renaissance} {\texttt{Renaissance~}}

\newcommand{\molH} {$\rm{H_2}$~}

\newcommand{\JU} {$\rm{ erg\ cm^{-2}\ s^{-1}\ Hz^{-1}\ sr^{-1}}$}
\newcommand{\JLW} {J$_{\rm LW}$}
\newcommand{\light} {\textit{light~}}
\newcommand{\heavy} {\textit{heavy~}}
\newcommand{\JHalo} {\texttt{1000J~}}
\newcommand{\Rapid} {\texttt{Rapid~}}
\newcommand{\PopIII} {\texttt{PopIII~}}
\def\jr#1{{\color{black} \bf(JR:  #1)}}

\begin{document}
\title{Massive Black Hole Seeds\vspace{-1.5cm}}
\author{John A. Regan$^1$}
\author{Marta Volonteri$^2$}
\thanks{$^*$E-mail:john.regan@mu.ie}
\affiliation{$^1$Centre for Astrophysics and Space Science Maynooth, Department of Physics, Maynooth University, Maynooth, Ireland \\
$^2$Institut d’Astrophysique de Paris, Sorbonne Université, CNRS, UMR 7095, 98 bis bd Arago, 75014 Paris, France}

\begin{abstract}
\noindent The pathway(s) to seeding the massive black holes (MBHs) that exist at the heart of galaxies in the present and distant Universe remains an unsolved problem. Here we categorise, describe and quantitatively discuss the formation pathways of both \light and \heavy seeds. We emphasise that the most recent computational models suggest that rather than a bimodal-like mass spectrum between \light and \heavy seeds with \light at one end and \heavy at the other that instead a continuum exists. \textit{Light} seeds being more ubiquitous and the heavier seeds becoming less and less abundant due the rarer environmental conditions required for their formation. We therefore examine the different mechanisms that give rise to different seed mass spectrums. We show how and why the mechanisms that produce the \textit{heaviest} seeds are also among the rarest events in the Universe and are hence extremely unlikely to be the seeds for the vast majority of the MBH population. We quantify, within the limits of the current large uncertainties in the seeding processes, the expected number densities of the seed mass spectrum. We argue that \light seeds must be at least $10^{3}$ to $10^{5}$ times more numerous than \heavy seeds to explain the MBH population as a whole.
Based on our current understanding of the seed population this makes \heavy seeds ($\rm{M_{seed}} > 10^3$ \msolarc) a significantly more likely pathway given that \heavy seeds have an abundance pattern than is close to and likely in excess of $10^{-4}$ compared to \light seeds. Finally, we examine the current state-of-the-art in numerical calculations and recent observations and plot a path forward for near-future advances in both domains. 
\end{abstract}
\maketitle

\begin{figure*}
\centerline{
  \includegraphics[width=17.0cm, height=10.0cm]{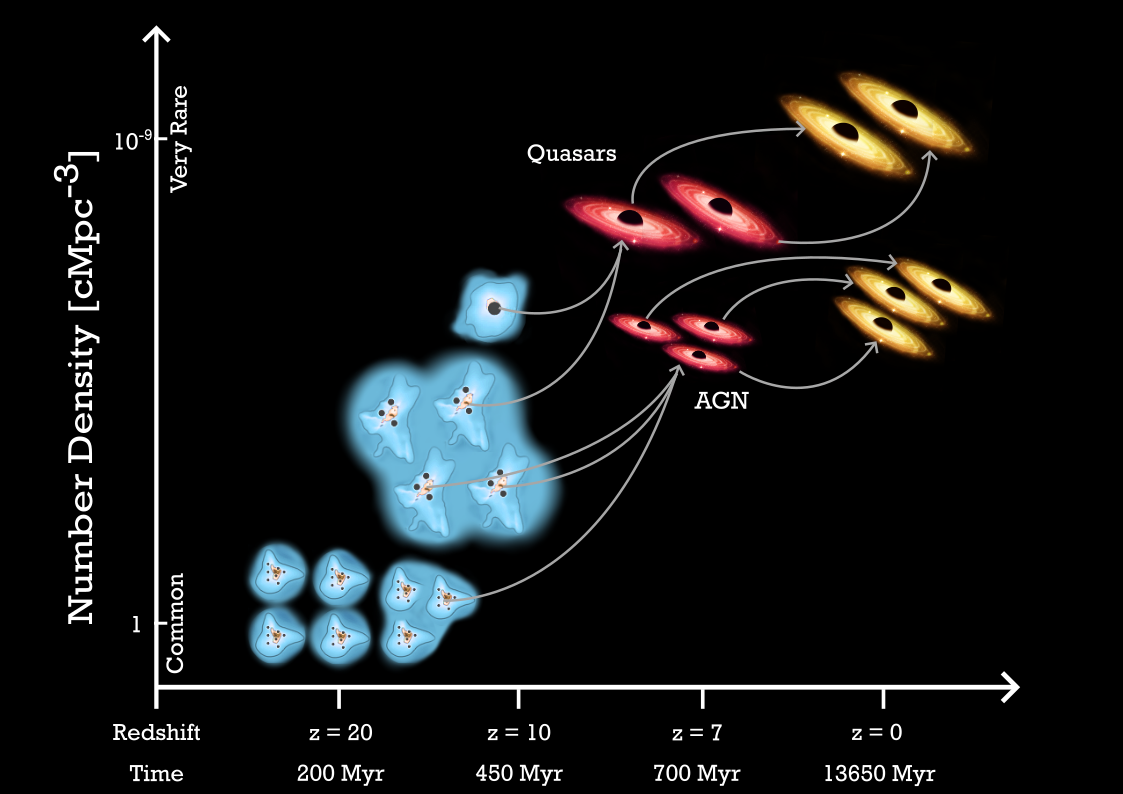}}
\caption{Schematic showing the Number Density (rarity, y-axis) versus Time and Redshift. \textit{Light} seeds can form in 'normal' galaxies with relative ubiquity in the early universe with number densities of the order of 1 cMpc$^{-3}$ at redshifts between z = 20 and z = 10. However, their (initially) small masses make them unlikely to be the progenitors for all of the `JWST' AGN or the high-z quasar population. On the other hand forming heavier seeds requires rarer environmental conditions and number densities varying from $10^{-1}$ to $10^{-5}$ cMpc$^{-3}$ (perhaps lower) but their initially heavier masses can facilitate more 
rapid growth potentially explaining the early MBH population as well as the high-z quasar population.}
\label{Fig:BHSchematic}
\end{figure*}

\section{Introduction} \label{Sec:Introduction}
\noindent Massive black holes (MBHs) with masses up to $10^{10}$ \msolar exist at the centre of most, perhaps all, massive galaxies. The ongoing challenge in the field is to understand the origin of the central MBHs. Up until recently the most distant MBHs 
discovered were quasars with masses up to $10^9$ \msolar observed at redshifts just beyond seven (approximately 700 Myr after the Big Bang) \citep{Fan_2006, Mortlock_2011, Venemans_2013, Venemans_2015, Banados_2018, Yang_2021, Wang_2021}. However, while these observations strongly indicate that the seeds for MBHs must form in the early Universe they offer little insight into their origin, as the initial conditions within which the seed black hole formed have long since been erased and the black hole itself contains no signature of that event. \\
\indent More recently, observations with JWST have uncovered MBHs with masses in the range $10^6$ \msolar to $10^8$ \msolar \citep{Kovacs_2024, Maiolino_2023, Larson_2023, Bogdan_2023, Matthee_2024, Greene_2024} at even higher redshifts. These MBH masses are lower than the high-z quasar masses previously detected but the galaxies within which these ``JWST-quasars'' have been detected likely represent the tip of the iceberg at this time in terms of both halo and stellar masses \citep{BoylanKolchin_2023, Juodzbalis_2024, Li_2024}. Nonetheless, with inferred masses down as low as $5 \times 10^6$ \msolar \citep{Maiolino_2023} these observations set an upper limit to the seed masses of this population of MBHs at the very least. The number density of these MBH host galaxies is currently unclear with potentially significant evolution of the central black hole following seeding. If we assume that current candidate AGN being observed by JWST are indeed true AGN then the number densities could be of the order $10^{-4}$ to $10^{-3}$  cMpc$^{-3}$ \citep{Matthee_2024,Greene_2024,Scholtz_2023}.\textit{The goal of this paper is to highlight what the current state-of-the-art in MBH seeding is, to identify challenges in the current models and to plot a path forward in terms of reconciling current theoretical models with observational data}. \\
\indent  The models for the astrophysical formation of MBH seeds can be loosely broken down into \light and \heavy seed models. Here, we define a \light seed as any seed black hole with a mass less than 1000 \msolar and a \heavy seed as any seed black hole with an initial mass in excess of 1000 \msolarc. A key challenge is then to understand what the mass spectrum and number densities of MBH seeds is and to understand the growth prospects of the seeds. We illustrate this challenge to the field in Figure 1.\\
\indent The crucial question for the black hole population is ``Does the entire black hole mass spectrum originate from stellar mass black holes with characteristic masses of approximately 40 \msolar and a tail to higher and lower masses, which grow through accretion and mergers''? This would imply a continuum from stellar black holes to MBHs, with a single origin for all of them. Or is another channel required to populate the MBH mass spectrum? Depending on the mass distribution of this channel then this channel will, if 
it exists, give rise to an additional peak in the mass spectrum of black holes. Tantalisingly this peak could, in theory, be detected by future gravitational wave observations \citep{Sesana_2007, Sesana_2009}. \\
\indent The simplest case is to assume that the seeds are indeed  \light and that they originated from the remnants of the first stars \citep[e.g.][]{Madau_2001}. However, significant challenges to \light seed growth exist and we will address this question in detail in \S \ref{Sec:LightSeeds}.\\
\indent Alternative routes to MBH formation may be through either a dynamical runaway pathway where either stellar collisions and/or black hole mergers result in a \heavy seed. Another possibility is a near monolithic collapse, or at least reduced fragmentation of a massive gas cloud, resulting in the formation of a supermassive star (SMS). Both of these mechanisms create a \heavy seed with masses in excess of approximately $10^3$ \msolar and in some cases possibly up to $10^5$ \msolarc. We will address these questions and gauge their possible number density in \S \ref{Sec:HeavySeeds}. We begin by first briefly reviewing the \light seed model and then the \heavy seed model. More in-depth reviews of the mechanisms can be found in reviews by \cite{Latif_Ferrera_2016, Woods_2017, Inayoshi_2020,Volonteri_2021}. Additional mechanisms, which we will not consider in this paper, include supermassive stars fuelled by dark matter annihilation \citep[e.g.][]{Spolyar_2008,Feng_2021, Singh_2023} or primordial black holes \citep[e.g.,][]{Carr_2018,Hasinger_2020}. We do not consider these mechanisms here as there is still considerable uncertainty in their formation pathway and their possible number densities. However, continued research into these mechanisms is warranted given the outstanding questions over the mainstream pathways, as we now discuss, as well as the ability of "non-standard" dark matter to potentially solve other dynamical hurdles \cite[e.g][]{Alonso_2024}.

\section{Light Seeds as the origin of all MBHs} \label{Sec:LightSeeds}
\noindent While
this is indeed the simplest explanation there are significant challenges in growing
\light seeds in order to explain the entire MBH mass spectrum. The problem is particularly acute when attempting to use \light seed growth to explain the appearance of MBHs at $z \gtrsim 7$. In order to achieve such growth the seed black holes would need to grow, uninterrupted, at the Eddington rate (under the assumption that \light seeds form from the first or second generation of stars). The Eddington accretion rate is given by 
\begin{equation}
  \rm{M(t) = M_{0}}\ \rm{exp} \Big( \frac{1 - \epsilon_{r}}{\epsilon_{r}}
  \frac{t}{t_{Edd}} \Big ),
\end{equation}
where $\rm{t_{Edd}} = 0.45 $ Gyr and $\epsilon_{r}$ is the radiative
efficiency. For a ``standard'' radiative efficiency of $\epsilon_{r} \sim
0.1$ and a black hole seed mass of $\rm{M_{0}} = 10^2$ \msolarc, it takes
nearly 1 Gyr to grow from $\rm{M(0)}$ up to $\sim 10^9$ \msolarc. While therefore in principle marginally allowable
with current observational constraints the growth must continue at the Eddington 
rate all the way up to the maximum mass (i.e. the duty cycle must be unity). 
Such a scenario has been shown to be exceedingly unlikely 
with numerous models and cosmological simulations showing that \light seeds do not grow efficiently and tend to remain 
at their seed masses across cosmic time \citep{Alvarez_2009, Milosavljevic_2009, Smith_2018, Spinoso_2023}. \\
\indent The reasons for this are complex and multifaceted but are primarily driven by the fact that the \light seed black holes are unable to sink to, or remain in, the galactic centre where gas densities are higher and also because they must compete for accretion with the star formation processes evolving nearby while local ionising radiation can further hamper accretion. However, in rare circumstances rapid accretion onto a \light seed may be possible \citep[e.g.][]{Alexander_2014, Zubovas_2021, Shi_2024}. We will return to this topic in \S \ref{Sec:Growth}. \\

\section{Heavy Seeds as a Sub-population} \label{Sec:HeavySeeds}
\noindent The need for \heavy seed black holes has largely been driven by observations of MBHs with masses up to and in excess of 
$10^9$ \msolar at redshifts beyond seven \citep[e.g.][]{Fan_2023}. With continued observations putting tight constraints on the \light seed pathway the strong possibility of \heavy seeds being the progenitors for the MBH which inhabit galactic centres strengthens \citep[e.g.][]{Bogdan_2023, Natarajan_2024}. \\
\indent A number of different avenues remain open to achieving \heavy seed black holes, including dynamical runaway processes, the direct collapse
of a SMS into a MBH and the direct formation of a MBH via galactic mergers. We now briefly discuss the environmental conditions 
required to drive each of these pathways but refer the interested reader to a review on the subject for a more in-depth analysis \citep[e.g.][]{Inayoshi_2020,Volonteri_2021}. \\
\indent Seminal work by \cite{PortegiesZwart_2004} showed that MBHs can be formed through runaway
collisions of stars in dense young star clusters. This work has been built on by several groups in more recent years with at times 
contrasting results \citep{Devecchi_2008, Katz_2015, Rizzuto_2021, Gonzalez_2021}. The differences can often be ascribed to different initial conditions and environmental factors. For example
\cite{Arca-Sedda_2023} has shown using detailed numerical models using the Nbody6+ +GPU code that in clusters of up to $10^6$ stars that 
black hole growth via stellar collisions tops out at a few hundred \msolarc. While on the other hand simulations by \cite{Reinoso_2023} in smaller clusters using the AMUSE \citep{Portegies_Zwart_2009, Pelupessy_2013} framework have demonstrated that stellar collisions can grow stars up to several tens of thousands of solar masses when accounting also for gas accretion. In any case, 
while convergence of results has perhaps not yet been achieved it does appear that initial black hole seed growth up to masses greater than 1000 \msolar does appear viable and hence this route to \heavy seeds appears plausible (although sensitive to cluster initial conditions). \\
Another related pathway is hierarchical BH mergers within a star cluster undergoing core collapse. In this case repeated black hole mergers can in principle grow a black hole up to masses in excess of 1000 \msolar \citep{Fragione_2018, Fragione_2022, Antonini_2019, Mapelli_2021, Arca-Sedda_2021}. As before however, the final black hole mass is sensitive to the conditions found in the cluster. Nonetheless, again a pathway to \heavy seed MBH does exist. \\
\indent In addition to the dynamical pathway \heavy seeds can form via the formation of a
very massive or super-massive star (SMS). In contrast to the formation of a dense cluster and subsequent dynamical runaway inside the cluster a (near-)
monolithic collapse is allowed to develop. In this case where fragmentation is suppressed, either through low metallicity or the lack of other efficient coolants, then
the possibility of forming very massive or SMSs arises. SMSs are defined by their extremely high accretion rate onto the stellar surface ($\rm{\dot{M}} \gtrsim 0.001$ \msolaryrc)
which, through the accretion of high entropy gas, causes the photosphere
of the star to expand  \citep[so long as the accretion rate is maintained,][]{Haemmerle_2018}. Such stars are expected to be
very red, in contrast to massive PopIII stars whose spectrum would be blue, and cool with effective temperatures of approximately $\rm{T_{eff}} \sim 5000$ K \citep{Hosokawa_2012, Hosokawa_2013, Haemmerle_2018, Woods_2021}.
With maximum theoretical masses of greater than $10^5$ \msolar \citep{Woods_2017} these SMS stars would be
ideal progenitors for MBHs and among the heaviest seeds possible.

\indent Finally, \cite{Mayer_2010, Mayer_2014, Mayer_2024} have postulated a "direct" mechanism for forming extremely massive black holes. In their model the merger of massive galaxies ($\rm{M_{Gal} \gtrsim 10^{11}}$ \msolarc) can result in the direct creation of a supermassive black hole with a mass up to $10^8$ \msolarc. The formation of the MBH directly bypasses both the stellar and seed formation stages and 
may explain the existence of the quasar population seen at $z \sim 6$. However, it cannot explain the existence of less massive 
black holes at high redshift as the mechanism relies on the merger of massive haloes ($\rm{M_{Halo}} \sim 10^{12}$ \msolar at z $\sim$ 10 corresponding to 4$\sigma$-5$\sigma$ peaks) to drive large gas inflows creating the central MBH. \\
\indent We note here that high resolution modelling including state-of-the-art physical models is crucial here. When attempting to model the collapse of gas creating the central MBH. resolution and/or physical models can result in fragmenting objects which are unrealistically massive. Early simulations of PopIII formation in particular suffered from this issue \cite[e.g.][]{Bromm_1999, Abel_2000} while more recent approaches (when modelling heavy seed formation) have shown, for example, the importance of resolving dynamical heating \cite[e.g.][]{Wise_2019} and ISM turbulence \cite[e.g.][]{Latif_2022}.
\begin{figure}
\centering
   \includegraphics[width=1\linewidth, height=7cm]{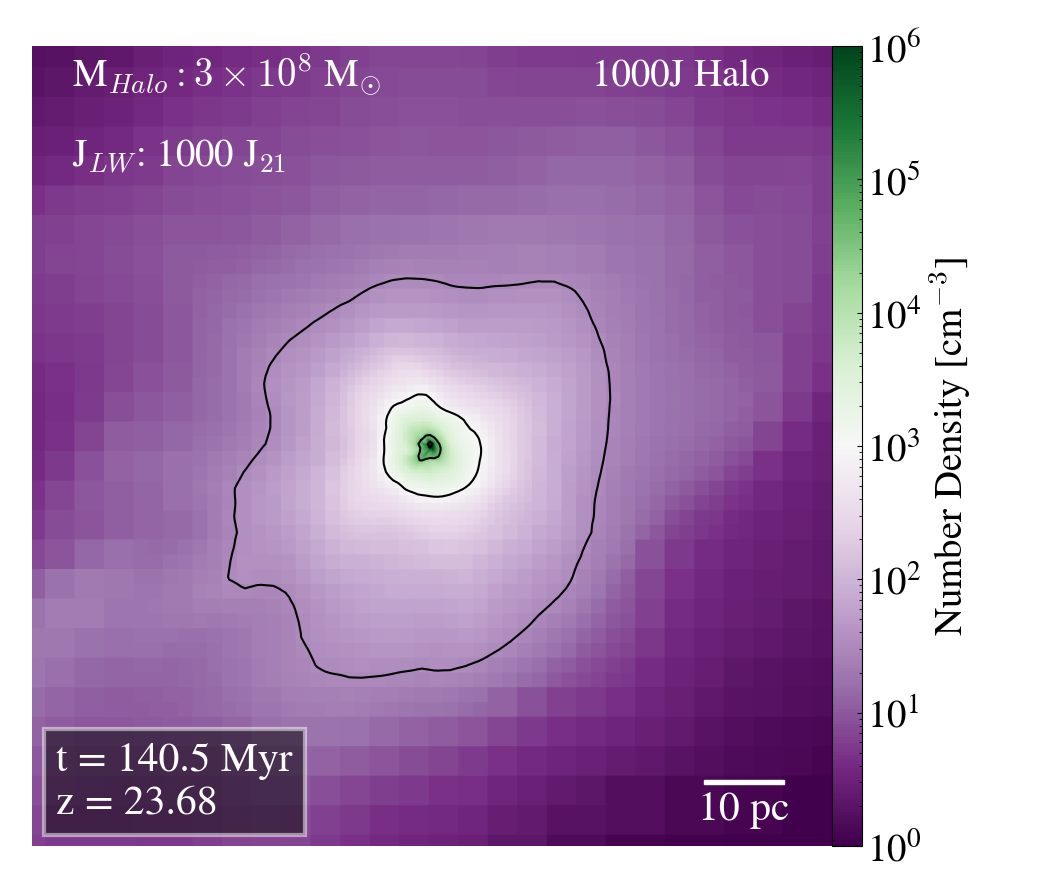}
   \includegraphics[width=1\linewidth, height=7cm]{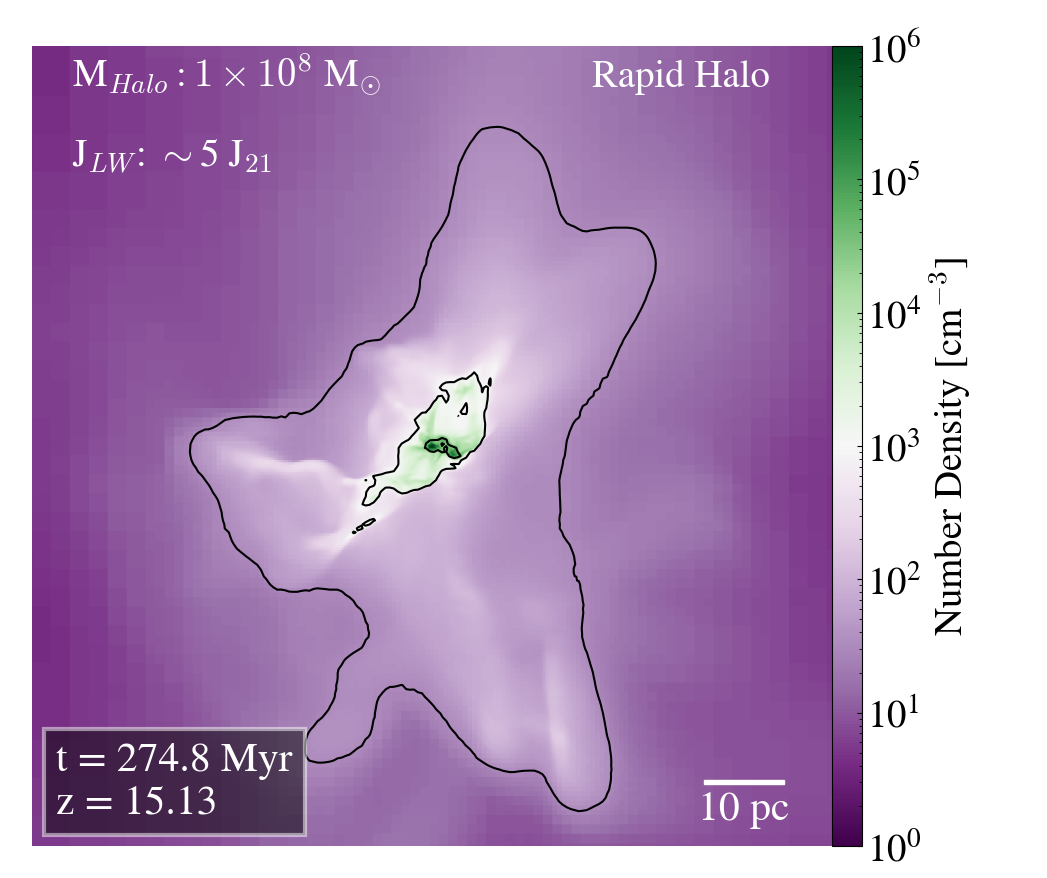}
   \includegraphics[width=1\linewidth, height=7cm]{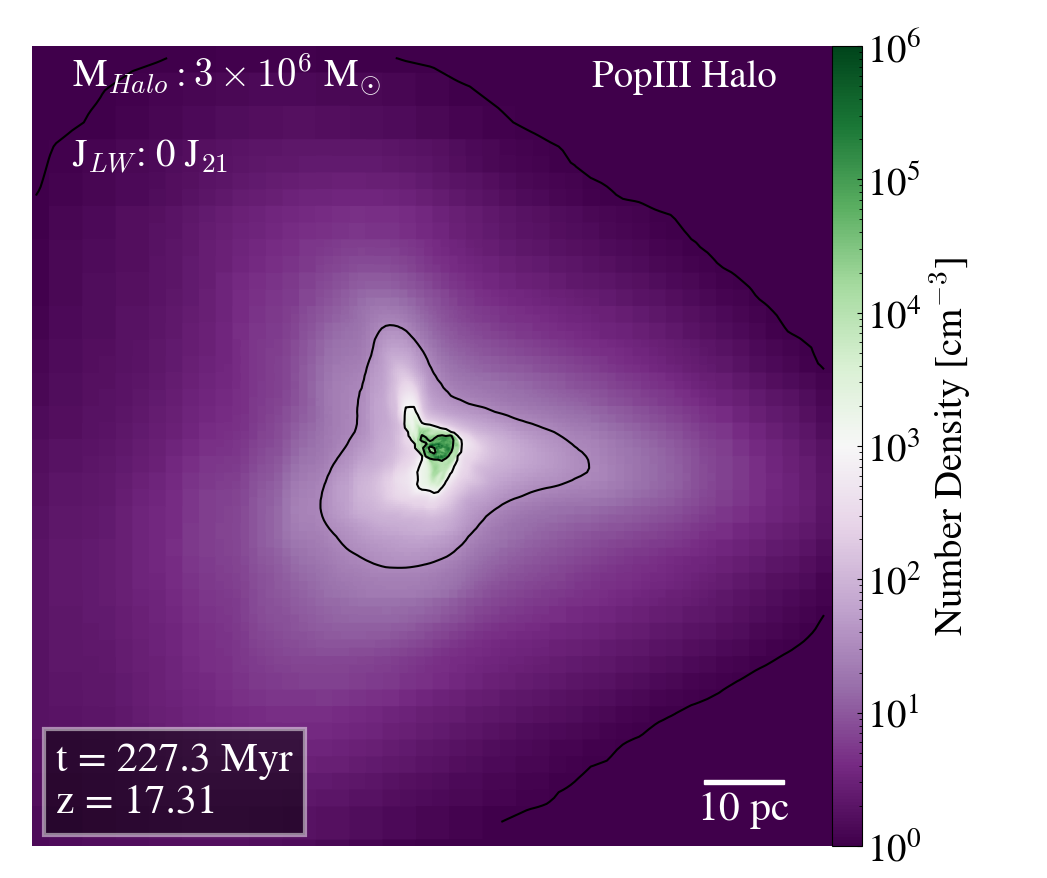}
   \caption{Gas Number Density projection plots from each of the three halos selected for our study. In the top panel we show the projection from the 1000J halo, in the middle panel the projection from the Rapid halo and in the bottom panel the PopIII halo.  Contours in each panel demarcate density levels in logspace. The 1000J halo (top panel) shows a very symmetric density profile, while both the Rapid (in particular) and PopIII haloes show much more diverse structure. Labelled in each panel is the halo mass, the LW flux impacting onto the halo, the redshift and time of the collapse. 
}
\label{Fig:DensityProjections}
\end{figure}

\section{Seed Number Densities}
\noindent A crucial question is what are the expected number densities of both \light and \heavy seeds in the high-z Universe. \textit{Light} seeds will be ubiquitous in the early universe - forming from the end points of massive stars. The exact number density of \light seeds is non-trivial to compute and will depend on the initial mass function of the first stars (which is unknown). However, a reasonable estimate would be between $10^{-1}$ cMpc$^{-3}$ and up to $10$ cMpc$^{-3}$ (at z $\sim 10$) \citep{Xu_2013, Smith_2018, Piana_2020, Trinca_2022}. The number density of \heavy seeds will of course be significantly less. \\ 
\indent By definition certain, perhaps rare, environmental conditions must occur for heavy seed formation pathways to be activated. On the contrary were such conditions common place, we would likely see them in action in the present day universe (at least for cases where gas/stellar metallicity does not need to be very small). Additionally, the rarity and perhaps uniqueness of this pathway to the early Universe has a significant impact on the expected number densities of \heavy seeds. See \cite{Woods_2017, Inayoshi_2020} for a review of \heavy seed formation environments. \\
\indent \cite{Dijkstra_2014} calculated the number density of heavy seeds forming through the so-called Lyman-Werner (LW) channel. In this case an atomic halo is irradiated with a super-critical LW by a nearby galaxy \citep{Dijkstra_2008, Agarwal_2012, Agarwal_2013, Visbal_2014b, Agarwal_2015b,Regan_2017}. In principle this can lead to the formation of an extremely massive seed (see \S \ref{Sec:ModelHaloes}). However, the calculated number densities are
small with expected number densities in the range $10^{-10}$ cMpc$^{-3}$ to $10^{-5}$ cMpc$^{-3}$ (at $z \sim 10$). It should be 
noted here that the higher number densities quoted here (i.e. $10^{-5}$ cMpc$^{-3}$) should be treated as strong upper limits as they rely on very optimistic formation criteria, namely, low values of the LW radiation needed to keep \molH dissociated or allowing for metal pollution.  Other authors have drawn similar conclusions on the expected number densities from this channel \citep{Inayoshi_2015b, Habouzit_2016, Agarwal_2015b,Trinca_2022}. \\
\indent Moving now to channels which potentially produce heavy seeds below the theoretical maximum mass (of SMSs) but well in excess of \light seeds masses. The so-called rapid assembly channel \citep{Yoshida_2003a, Fernandez_2014, Wise_2019, Regan_2020b, Lupi_2021, Latif_2022} can produce \heavy seeds with number densities in the range 10$^{-5}$ to $10^{-1}$ cMpc$^{-3}$ (at $z \sim 10$) (McCaffrey et al. in prep)  depending on environmental factors. Number densities achieved through other channels like baryonic streaming velocities result in number densities close to approximately 10$^{-5}$ cMpc$^{-3}$ (at $z \sim 10$) as well \citep{Hirano_2017}. \\
\indent Overall the number densities of \heavy seeds range over several orders of magnitude (varying according to the expected mass of the seed and the environmental conditions required) from $10^{-10}$ cMpc$^{-3}$ up to $10^{-1}$ cMpc$^{-3}$ (at $z \sim 10$).
The variance of these estimates spanning almost 10 orders of magnitude speaks to the complexity of the problem. However, what should be kept in mind here is that while producing the heaviest seeds may be exceedingly rare - producing more moderate mass seeds becomes progressively more straightforward and the higher number densities. This leads to the conclusion that even in the case of a \heavy seed channel that the seed mass function could well be a continuum with a peak at some characteristic mass. We now move on to discuss the types of haloes expected to form both \light and \heavy seeds.

\begin{table*}[!t]
  \centering
  \begin{tabular}{ |p{3cm}|p{2.5cm}|p{1.5cm}|p{3cm}|p{4.0cm}| } 
    \hline
    \textbf{Model Name} & \textbf{Halo Mass [\msolarc]} & \textbf{Redshift} & \textbf{Mass of Most Massive Star [\msolarc]} & \textbf{Halo Number Density [cMpc$^{-3}$] at $z \sim 10$} \\
    \hline
    \JHalo Halo & $2.59 \times 10^7$ & 23.70 & 76123 & $10^{-10} - 10^{-5}$ \\
    \hline
    \Rapid Halo & $9.3 \times 10^7$ & 15.05 & 6127 & $10^{-5} - 10^{-1}$ \\
    \hline
    \PopIII Halo & $3.7 \times 10^6$ & 17.21 & 173 & $10^{-1} - 10^{+1}$ \\
    \hline
   
  \end{tabular}
  \caption{\noindent Characteristic of the three different haloes used to illustrate our arguments. The three haloes are the \JHalo halo, the \Rapid halo and the \PopIII halo. For each halo we give the halo mass (second column), the redshift at which star formation first occurs in each halo (third column), the mass of the most massive star in each halo (fourth column) and in the last column the expected number density of each halo type.} \label{Table:HaloStats}
\end{table*}

\begin{figure}
\centering
  \includegraphics[height=7.0cm]{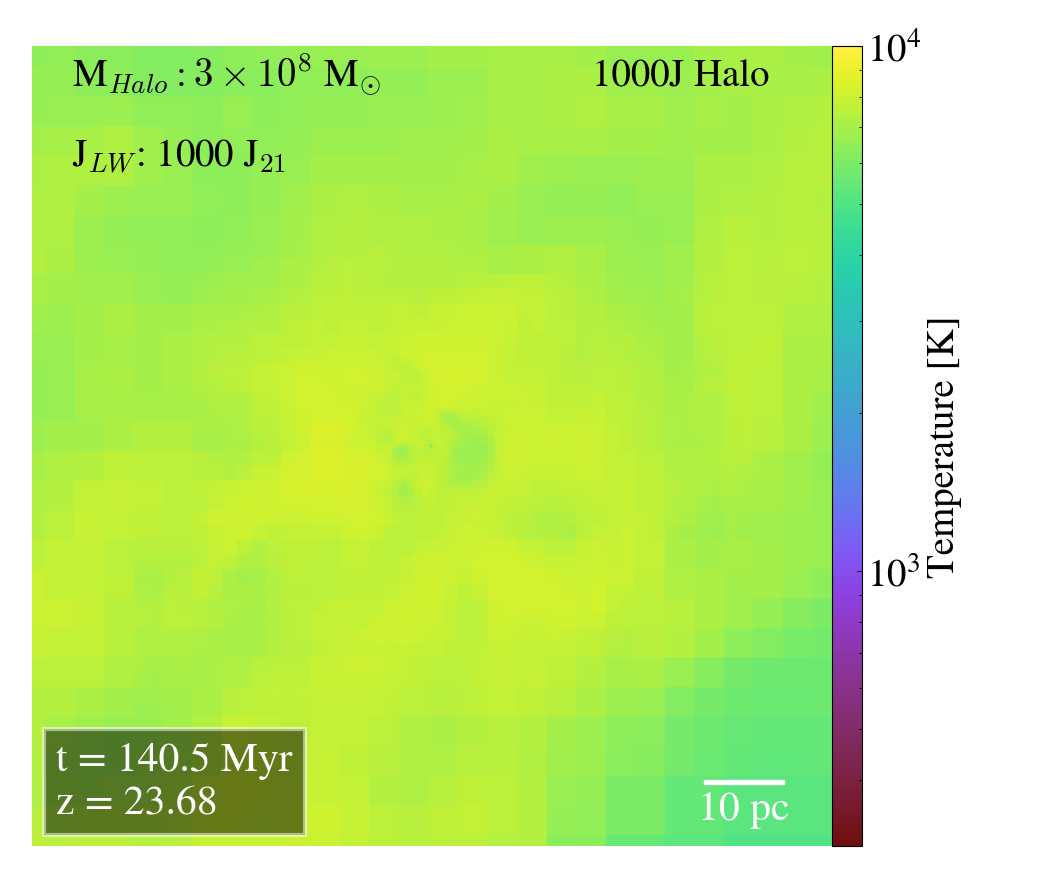}
  \includegraphics[height=7.0cm]{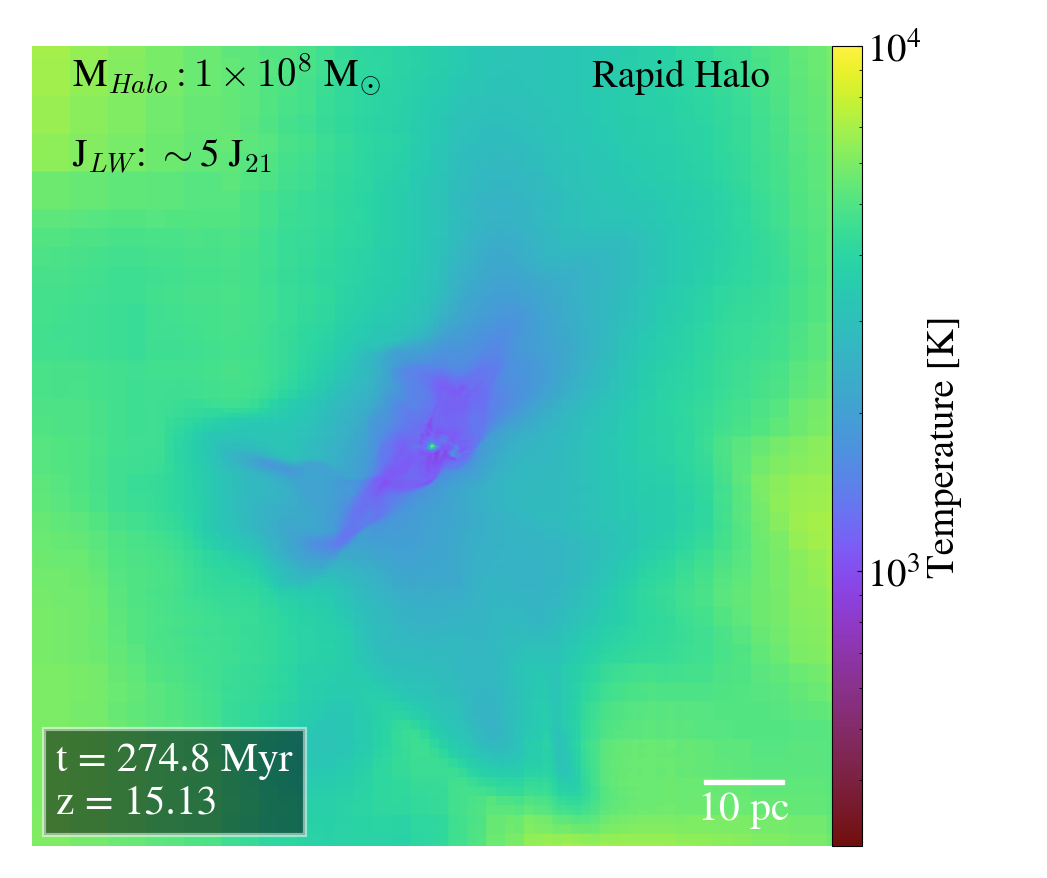}
    \includegraphics[height=7.0cm]{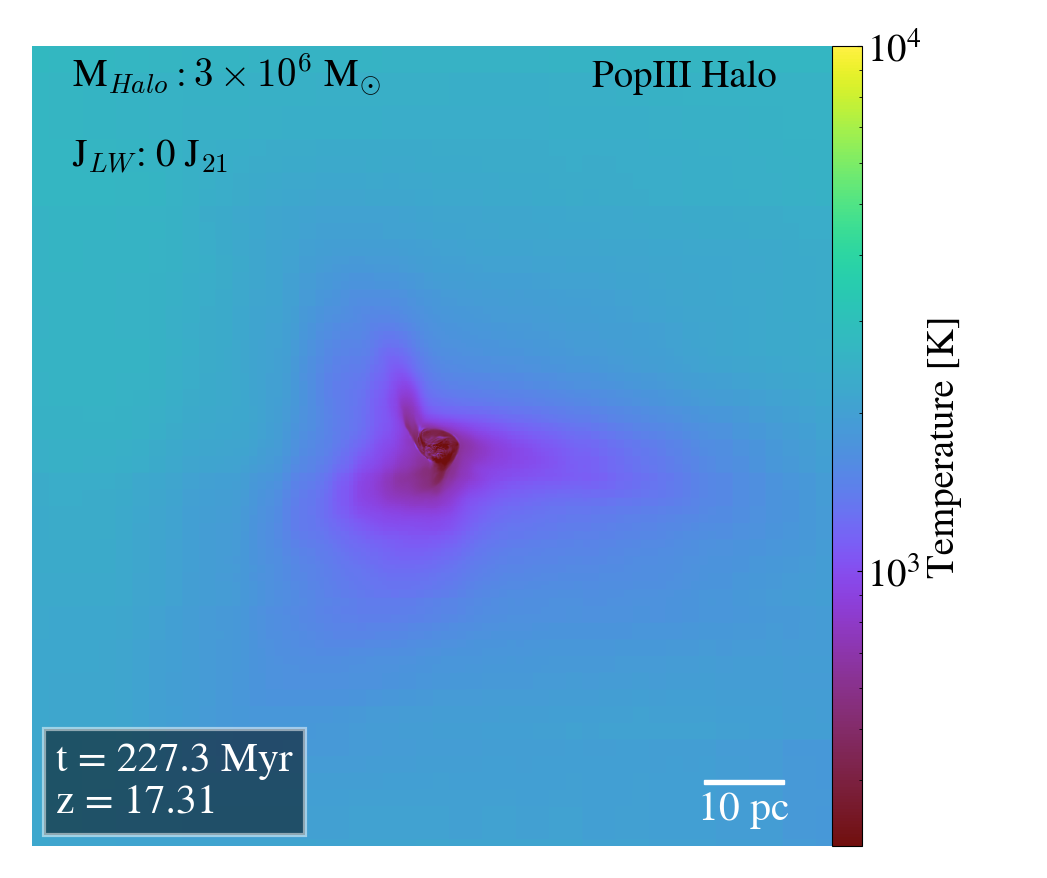}   
\caption{Similar to Figure \ref{Fig:DensityProjections} except using the temperature field as the projection variable. In the top panel we show the projection from the 1000J halo, 
in the middle panel the projection from the Rapid halo and in the bottom panel the 
PopIII halo. The 1000J shows very little temperature differences reflecting the fact that the strong LW field ($\rm{J_{LW} = 1000 \ J_{21}}$) largely dissociates the \molH ensuring the collapse is isothermal. The middle panel, with only a very mild LW field ($\rm{J_{LW} \sim 5 \ J_{21}}$), shows significantly more temperature structure 
(which drives more structure in the density field). Likewise the bottom panel shows a strong temperature gradient as \molH is able to cool the gas to below 1000 K in the halo centre. 
}
\label{Fig:TemperatureProjections}
\end{figure}

\section{Model Haloes} \label{Sec:ModelHaloes}
\noindent To illustrate the different environmental conditions that have been investigated in the literature over the last decade or more we have selected three haloes that broadly reflect the mainstream pathways of investigation. Additionally, the environmental conditions required feed directly into the expected number densities of these seeds. We are not aiming for complete sampling of all scenarios and environments investigated but rather the small sample of haloes selected here reflects in broad terms the emergence of \light and \heavy seed pathways. The haloes we select we name the \texttt{Rapid} halo, the \JHalo halo and the \PopIII halo. The characteristics of each halo are given in Table \ref{Table:HaloStats}. \\
\indent The \JHalo halo is the prototypical halo used to test the intense LW field model (see \S \ref{Sec:HeavySeeds} for details). The \Rapid halo is the result of the rapid assembly paradigm and results in the formation of \heavy seed black holes via the rapid assembly (or dynamical heating) mechanism. Finally, the \PopIII halo is a standard mini-halo whose environment is not influenced by either (extreme) LW radiation or rapid assembly. The \PopIII halo gives rise to \light seeds while the \JHalo halo and \Rapid halo give rise to \heavy seeds albeit almost certainly of different initial masses. The expected number densities of seeds as well as their potential mass ranges are given in Table 1. We now discuss each of the haloes in detail comparing and contrasting the different environmental conditions and the reasons why different halo configurations lead to \light or \heavy seed formation and whether they offer the potential for significant future growth.

\subsection{The \JHalo Halo}
\noindent The \JHalo is taken from \cite{Regan_2018b} and consists of a single halo irradiated by a 
LW background field with flux 1000 J$_{21}$\footnote{J$_{21}$ is in units of $10^{-21}$ \JU} (an intensity of 1000 J$_{21}$ is generally accepted in the literature as required for substantial \molH suppression \citep[e.g.][]{Shang_2010}). The simulation evolves with this background field strength until one halo exceeds the atomic cooling threshold. At this point 
cooling due to Lyman-alpha line emission dominates the cooling and allows the gas to collapse almost isothermally at approximately 8000 K. The most massive star formed in the halo has a
mass of approximately 76,000 \msolar and there were only four stars in the halo at the end of the simulations ($\sim 250$ kyr after the formation of the first star). This is the proto-typical 
isothermal collapse seen by most studies in which collapse in the presence of a large LW field is studied \citep{Latif_2013d, Visbal_2014c,Latif_2016, Dunn_2018, Schauer_2021}. Even larger LW fields may result in less fragmentation at the scales studied while lower LW fields will likely result in more fragmentation (taking into account halo stochasticity). \\
\indent In the top panel of Figure \ref{Fig:DensityProjections} we show a gas number density projection centred on the \JHalo halo. The halo is highly symmetric indicating a near spherical collapse. The symmetry of the 
halo is also evidenced by the (density-weighted) temperature projection plot shown in the top panel of Figure \ref{Fig:TemperatureProjections}. The strong LW field homogenises the temperature field leading to a near isothermal collapse (at $\sim 8000 \rm{K}$). The homogenisation of the temperature field has a knock on effect on the density - making it more homogeneous as well due to the removal of temperature gradients which can lead to cooling instabilities. This in turn leads to less fragmentation. This is an ideal 
scenario for creating heavy seeds. \\
\indent In this specific case most of the stellar mass ends up in two SMSs with four SMSs forming in total. The mass of each of the stars in shown in Figure \ref{Fig:Masses}. The accretion rate onto the most massive star in the \JHalo halo is shown in the top panel of Figure \ref{Fig:Accretion} (green line). The SMS accretes at super-critical rates \citep{Nandal_2023} for more than 200 kyr before the gas supply becomes restricted. The final mass of each of the stars are shown in Figure \ref{Fig:Masses} (green bars). In Figure \ref{Fig:Accretion} we show the mass evolution and accretion rate onto the most massive star in the simulation (green lines). The most massive star achieves a mass of more than 76,000 \msolar after approximately 250 kyr. At this point the accretion rate falls off as most of the locally available gas is consumed and the star contracts to the main sequence. \\
\indent The number density of haloes subjected to a super-critical LW flux cannot be obtained from the simulations 
discussed above due to their idealistic nature. However, both \cite{Visbal_2014b} and \cite{Regan_2020} (using the \renaissance simulation suite) computed the abundance of so-called synchronised pairs which can lead to the emergence of haloes similar to the \JHalo halo \citep{Regan_2017}. Both authors found the abundance of synchronised haloes to be less than $10^{-2}$ cMpc$^{-3}$ (at z $\sim 10$). However, this number should be seen as an upper limit. For the "Normal" region of the \renaissance simulation suite \cite{Regan_2020} found exactly zero synchronised pairs while finding five, in the clustered, "Rarepeak" region. Taking into account the relative rarity of the Rarepeak region, the number densities obtained should be decreased by a factor of approximately 1000. This leads to number densities of less than $10^{-5}$ cMpc$^{-3}$ (at z $\sim 10$) for a \JHalo like object. An important additional point here is that in-situ LW radiation can potentially ease the thresholds somewhat but remains in delicate balance with other factors like metal-pollution which can suppress heavy seed formation \cite[e.g.][]{Dunn_2018, Bhowmick_2022a, Chiaki_2023}.

\subsection{The \Rapid Halo}
\begin{figure}
\centering
\includegraphics[width=8.0cm, height=6.0cm]{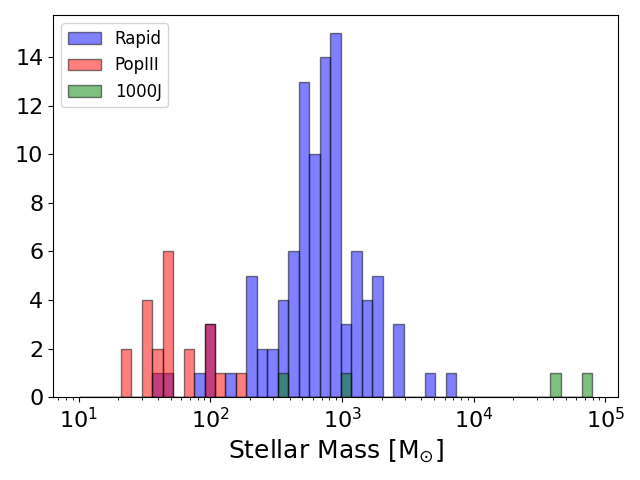}
\caption{The masses of all of the stars formed in each of the three characteristic haloes. Only four stars form in total in the \JHalo halo after 250 kyr, approximately 100 in the \Rapid halo after 2 Myr while the \PopIII halo forms approximately 20 stars after 2 Myr. The masses of the stars formed in the \JHalo are up to an order of magnitude more massive than those in the \Rapid halo and more than two orders of magnitude more massive than those formed in the \PopIII halo. 
}
\label{Fig:Masses}
\end{figure}

\noindent The \Rapid halo is taken from \cite{Regan_2020b} which were themselves zoom-in simulations of the \renaissance simulation suite. The halo was originally discovered as part of \cite{Wise_2019}. In this paradigm, as described in \S \ref{Sec:HeavySeeds}, the halo experiences periods of extremely rapid growth, which dynamically heats the halo, overcoming the ability of \molH to cool the gas. As a result the gas stays `hot' increasing the Jeans mass of any fragmenting clumps (and reducing the number of fragmenting clumps) - see also \cite{Latif_2022} who explore a similar case to that of \cite{Wise_2019} albeit in a rarer environment. \\
\indent The gas number density distribution of the halo is shown in the middle panel of Figure \ref{Fig:DensityProjections}. The morphology is clearly very different to that of the \JHalo shown in the panel above. In the case of the \Rapid halo there is a distinct lack of symmetry. The (density-weighted) temperature projection shows a similar diversity compared to the \JHalo halo. The temperature distribution with cooler gas towards the centre clearly contains far more structure. The \Rapid halo forms 99 stars over the course of 2 million years with the most massive star having a mass of 6127 \msolar at the end of the simulation. The mass of each star at the end of the simulation is shown in Figure \ref{Fig:Masses} (blue bars). The halo has a peak star mass of approximately 1000 \msolarc. The mass of the most massive star and the accretion rate onto the  most massive star in the \Rapid halo is shown, as a function of time, in Figure \ref{Fig:Accretion} (blue lines). Similarly, to the \JHalo the most massive star stops accreting after a few tens of kyr as the locally available gas supply runs out. At this point the star contracts to the main sequence and becomes a massive (and extremely luminous) PopIII star.\\
\indent The number density of rapidly growing haloes from the \renaissance simulations is approximately 
$5 \times 10^{-3}$ cMpc$^{-3}$ at z $\sim 10$ \citep{Regan_2020a}\footnote{This value is taken from the 'Normal' simulation which runs to z = 11.6}. This is at least two orders of magnitude higher than the probablility of forming a \JHalo like object. The criteria used to define these haloes from the \renaissance simulations was all haloes with mass inflow rates exceeding 0.1 \msolaryr in the halo centre and with metallicities $Z \lesssim 10^{-3}$ Z$_{\odot}$.

\subsection{The \PopIII Halo}
\noindent As with the \Rapid halo the \PopIII Halo is taken from \cite{Regan_2020b}. The halo represents a canonical mini-halo with no LW field flux or rapid assembly. The \PopIII halo forms PopIII stars out of primordial Hydrogen and Helium. The bottom panel of Figure \ref{Fig:DensityProjections} shows the gas number density projection of the collapsing \PopIII halo. The collapse is somewhat symmetric while also showing clear evidence of an extended morphology. Similarly,  the (density-weighted) temperature projection shown in the bottom panel of Figure \ref{Fig:TemperatureProjections} shows a clear temperature gradient with cooler gas forming deeper inside the halo. The \PopIII halo forms 21 stars by the end of the simulation with the most massive star having a mass of 173 \msolarc (see Figure \ref{Fig:Masses}, red lines). The top panel of Figure \ref{Fig:Accretion} shows the cumulative mass as a function of time for the most massive star (red line) while the bottom panel shows the accretion rate onto the most massive star in the \PopIII halo. \\
\indent The number density of haloes, at z $\sim 10$, in the Renaissance simulation suite hosting \light seeds (i.e. PopIII remnant black holes) is approximately 0.3 cMpc$^{-3}$. This is approximately $10^3$ times higher than the number density of rapidly accreting haloes from the same simulation volume and at least $10^5$ times higher than the number density of LW haloes expected in a similar volume. 

\section{Model Halo Comparison} \label{Sec:HaloCompare}
\begin{figure}
\centering
  \includegraphics[height=7.0cm]{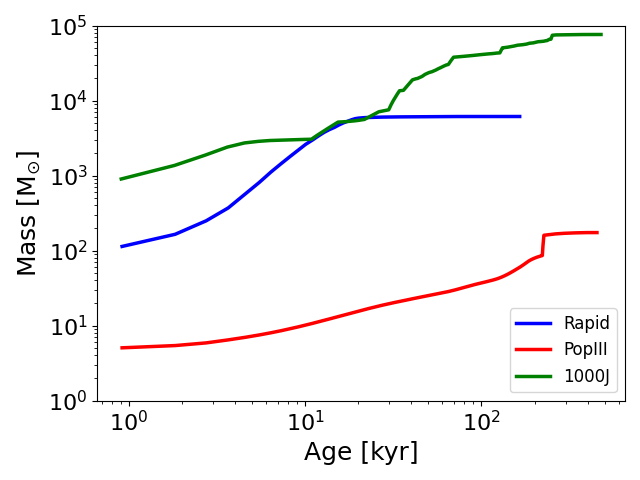}
  \includegraphics[height=7.0cm]{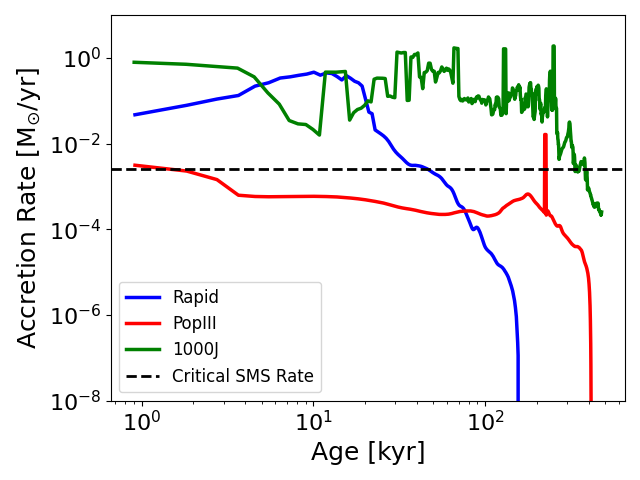}
\caption{The evolution of the most massive star that forms in each of the three haloes (1000J, Rapid and PopIII). Stars are modelled as accreting sink particles. The spatial resolution of each simulation is approximately $10^{-3}$ pc (physical). In the top panel we show the mass as a function of time. In the bottom panel we show the accretion rate onto the most massive star in each halo as a function of time. The \JHalo halo forms the most massive object followed by the \Rapid halo and the \PopIII halo. The dashed line shows the 
critical accretion that must be maintained onto the stellar surface to meet the criteria for being a super-massive star \citep{Nandal_2023}.
}
\label{Fig:Accretion}
\end{figure}

\noindent Differences are clearly apparent across the three characteristic haloes chosen.
The \JHalo has a near homogeneous density distribution as a result of the 
super-critical LW flux it is exposed to. This density (and temperature) homogeneity  promotes a strong suppression of fragmentation resulting in two extremely massive objects forming (masses of 76000 \msolar and 44000 \msolarc) along with two less massive stars. The other two haloes (\Rapid and \texttt{PopIII}) show a more inhomogeneous density (and temperature) distribution (due to the presence of higher \molH fractions) and result in lower mass stars. The distribution in stellar masses across each halo is shown in Figure \ref{Fig:Masses} with the \JHalo showing fewer but significantly more massive stars compared to either the \Rapid halo or the \PopIII halo. \\
\indent There are two primary reasons for the elevated stellar masses found in the \JHalo compared to the \Rapid and \PopIII halo. On the one hand and as described above the LW flux dissociates \molH and leads to larger Jeans masses and hence larger protostars \citep{Prole_2024} but also these protostars can accrete more readily and for longer (see Figure \ref{Fig:Accretion}). In the \Rapid halo and \PopIII halo the structures are far more asymmetric making continued accretion much more difficult (at least for the stellar masses probed here). For the cases of stars in both the \Rapid and \PopIII haloes the stars 
can initially accrete gas readily but the gas supply quickly gets consumed and the stars contract to the main sequence \citep{Regan_2020b} with masses significantly below that seen in more symmetric and homogeneous halos.\\
\indent It should be noted that in Figure \ref{Fig:Accretion} we show only accretion over a relatively short period (hundreds of kilo-years). The time-resolution needed for these simulations is very short - of the order of years compared to large scale cosmological simulations with timesteps many orders of magnitude higher.
The subsequent growth of the seed (\light or \textit{heavy}) will depend fundamentally on the evolution of the host galaxy over several tens and hundreds of megayears. Presently, simulations able to track the model the formation of  \light and \heavy seeds and to subsequently follow their growth are extremely computationally challenging and essentially intractable. We nonetheless describe progress in this direction in the following section bearing in mind that this remains an extremely challenging modelling problem. \\

\section{Black Hole Growth} \label{Sec:Growth}
\noindent The sustained growth of black holes is clearly critical for their evolution into the MBHs we observe as both high-z quasars and as the "JWST-quasars" at z $\gtrsim 7$. The question then becomes which seeds are more likely to experience significant growth? 

\subsection{Growth of Light Seeds}
\noindent Light seeds can suffer from initially severely stuttered growth as a result of their formation pathway. Depending on their final mass the seeds are often born "starving" into an environment which is hostile to initial growth \citep[e.g.][]{Whalen_2004, Alvarez_2009, Milosavljevic_2009}. This is because the ionising feedback from the  stellar phase plus the energy injected into the surrounding gas results in a \light seed born into a relatively underdense local environment.\\
\indent Light seeds can nonetheless grow in two ways, firstly via gas accretion and secondly via mergers with other compact objects. 
Due to their smaller masses \light seeds have a significantly smaller Bondi radius (which scales as M$^2_{BH}$) compared to \heavy seeds. Furthermore, their mass relative to that of their host galaxy and indeed the surrounding stellar and gas distribution is negligible. This means that \light seed black holes will tend to move easily within their environment - a movement which can appear random. A number of authors have investigated this property of \light seeds numerically. \\
\indent \cite{Smith_2018} et al. used the \renaissance suite 
to track, in post-processing, the evolution of \light seed black holes formed from PopIII stars. They tracked the 
evolution of over 20,000 \light seeds for up to 200 Myrs and found negligible growth across the entire 
\light seed population. The reason being is that the \light seeds experience extremely low duty cycles due to the surrounding (negative) feedback processes of star formation and stellar feedback. Moreover, higher resolution studies of (PopIII) star forming environments show that PopIII stars form in multiples \citep[e.g.][]{Turk_2009, Clark_2011a, Prole_2024} setting off a competitive accretion scenario further mitigating against efficient accretion. Other authors \citep[e.g.][]{Pfister_2019, Ma_2021}  find that only seeds with masses in the \heavy seed (up to $10^8$ \msolar in the case of the \cite{Ma_2021} study) range can effectively sink to the centre of the potential well where efficient accretion may be sustainable. None of the above studies rule out \light seed accretion but instead find that the probability of efficient \light seed growth is extremely low. \\
\indent Building on these challenging conclusions from studies of \light seed growth a number of groups have investigated how growth can be accelerated when the conditions are favourable. In the majority of these cases the 
\light seed black hole is modelled to either sit at the centre of a converging flow \citep[e.g.][]{Volonteri_2005b} or is "trapped" within the confines of a nuclear star cluster \citep{Alexander_2014, Natarajan_2020} or within a circumnuclear disk \citep[e.g.][]{Lupi_2016}. More recent work by \cite{Shi_2024} shows that a small fraction ($\sim$ 1\%) of \light seeds can achieve bursts of rapid growth 
when placed in environments that are favourable (i.e. are gas rich and support growth).
In each of these cases the growth of the initially \light seed proceeds via bursts of super-Eddington accretion. The physical process of super-Eddington accretion has been demonstrated extensively via extremely high resolution simulations \citep[e.g.][]{Sadowski_2009, Sadowski_2016, Sadowski_2016a, Jiang_2019, Jiang_2020}. However, what is lacking are robust numerical simulations which show 
self-consistently the growth of \light seeds in a cosmological setting. The reason for this is that a combination of high resolution, with resolved dynamics, and a realisation containing a rare dynamical occurrence is required. This has, thus far, not been realised. However, such a setup should be tractable with current simulation codes and techniques at least for the early evolutionary phase. \\
\indent In the above discussion we have focused on growth through gas accretion. However, a \light seed can grow 
also via mergers. The classic case here is a dense stellar cluster where stars undergo collisions resulting in \light or \heavy seed formation and the seed subsequently merges with other stars and BHs resulting in a final seed mass. The final seed mass attainable here is 
unclear and detailed state-of-the-art n-body simulations have not yet converged on a final result \citep[e.g.][]{Arca-Sedda_2023, Rantala_2024} due to the complex parameter space that must be explored. Additionally, the impact of gas on both the dynamics and accretion 
must be incorporated into this scenario and this could potentially result in much larger seed masses \citep[e.g.][]{Chon_2020, Reinoso_2023}. \\
\indent Conversely, \light seed mergers in the aftermath of galaxy mergers (the classic case considered, for instance, to estimate merger rates for gravitational wave detectors such as LISA) is unlikely. This is because of the same reasons why sinking and retaining \light seeds in galaxy centres is difficult, as already discussed. In a galaxy merger a \light seed will have mass not much larger than that of stars and therefore dynamical friction is inefficient: the probability that two \light seeds, which have very small cross section, find each other within this sea of stars is exceedingly small. 

\subsection{Heavy Seed Growth}
\noindent Given that the growth of \light seeds appears very inefficient in the main, it is worth exploring whether \heavy seeds can avoid this bottleneck. As discussed above this will likely depend sensitively on the mass of the \heavy seed. \cite{Pfister_2017} showed that growth is extremely inefficient below at least $10^5$ \msolar and 
perhaps as high as an initial mass of $10^8$ \msolar in the case of \cite{Ma_2021}. These seminal works appear to be borne out by more recent simulations. For example \cite{Lupi_2024} model the progenitor of a 
high-z quasar using an initial black hole seed mass of $10^5$ \msolar (the  model sets the initial dynamical mass of the 
\heavy seed to be $10^6$ \msolarc). This initial mass helps to stabilise the seed prior to rapid growth - nonetheless the initial growth is stunted by supernova feedback and requires significant halo growth to facilitate the black hole growth \cite[see also][]{Bhowmick_2022}. Although the initial barrier to growth depends on the strength of supernova feedback, what this shows is that growth can occur onto a \heavy seed but only if the seed is massive \underline{and} if the conditions are favourable (i.e. the halo is sufficiently massive). \cite{Latif_2020} reached similar conclusions, albeit from somewhat more idealised conditions - following the growth of a \heavy seed (M$\rm{_{seed}} = 10^5$ \msolarc) starting from a halo similar to the \JHalo halo described above. 
\\\indent On larger scales the Astrid \citep[e.g.][]{Bird_2022, Ni_2022}, NewHorizon \citep{Dubois_2021, Beckmann_2022} and Romulus \citep[e.g.][]{Tremmel_2017, Ricarte_2019} simulation suites all seed black holes with masses well inside the \heavy seed regime as well as implementing dynamical friction subgrid modelling. For higher mass galaxies the results for the simulatiion suites are boardly consistent. However, for smaller mass galaxies (M$_* \lesssim 10^9$ \msolarc), the results are not converged. \cite{Ricarte_2019} show that MBHs can grow in galaxies independent of the stellar mass (and hence show little or no effect from supernova feedback). This 
is in contrast to other results \citep[e.g.][]{Dubois_2014, Habouzit_2017, AnglesAlcazar_2017}. The reasons for this difference are not obvious and require further investigation \citep{Ricarte_2019}.\\
\indent So overall where does the extensive work on black hole growth leave us? Across the various simulation setups
there is convergence that black holes can grow once the host halo becomes massive enough with some differences arising for black hole 
growth in lower mass galaxies which needs to be fully understood. Additionally, the black hole must achieve a mass of at least $10^5$ \msolar before growth can be achieved - this is particularly evident from the Astrid simulations which have the lowest seed masses 
($\rm{M_{seed} \gtrsim 10^4}$ \msolarc) and see very limited growth for the lightest seeds. Similar results are obtained in semi-analytical models \citep{Trinca_2022,Spinoso_2023}. The main result therefore is that any MBH seeds below approximately $10^5$ \msolar grow extremely inefficiently due to dynamics. In addition to this, in low mass galaxies ($\rm{M_* \lesssim 10^9}$ \msolar) growth may by hampered by supernova feedback but this result suffers from different interpretations in the literature and requires further study.  \\

\section{Summary and Conclusions} \label{Sec:Conclusions}
\noindent  The goal of this paper is to weave together recent developments in modeling MBH seeding and pathways to their growth. The current paradigm is broken into \light and \heavy seeds with \light seeds typically defined as those with initial masses less than 1000 \msolar and \heavy seeds as those with initial masses in excess of 1000 \msolarc \citep[e.g.][]{Sassano_2021, Evans_2023, Jeon_2024}. Scenarios considered in the past gave rise to a bimodal initial mass function:  common \light seeds related to PopIII stars with mass $\sim 100 $ \msolar and rarer \heavy seeds related to SMSs with mass $\sim 10^5 $ \msolarc. In order to grow these seeds into the massive and supermassive regime, two main pathways were considered: super-Eddington on  \light seeds and sub-Eddington on \heavy seeds, although of course super-Eddington accretion can occur, and indeed it may be more likely to occur, for \heavy seeds.  What the most recent simulations are instead suggesting is that these two regimes are likely to be the extreme cases of a broad distribution of initial properties. 

To characterise the pathways to achieving each seed (\light or \textit{heavy}) we select three representative haloes which we use to illustrate the different environmental conditions in which they form. Our findings regarding seed \textit{formation} are as follows:

\begin{enumerate}

    \item \textbf{The heaviest seeds:} Haloes subjected to a strong or super-critical LW flux will most likely result in the largest single stellar object forming, up to $\sim 10^5 $ \msolar (\JHalo halo). This is because the absence of any coolant, apart from Hydrogen, results in a halo whose temperature is nearly uniform across the halo inducing a near spherical collapse. Such conditions are very rare (seed number densities $\sim 10^{-10} -10^{-5} \ \rm{cMpc}^{-3}$ at z $\sim$ 10) and cannot explain the number densities of either the high-z AGN population or the local MBH population.
        \item \textbf{Heavy seeds:} Other mechanisms that lead to environments conducive to heavy seed formation (e.g. a rapid assembly process, baryonic streaming velocities or mechanisms that lead to the emergence of dense stellar systems) but not necessarily a monolithic collapse will 
        generate a spectrum of heavy seed masses. 
        In this, more fragmented, environment heavy seed multiplicity will be higher with the mass therefore spread across more fragments. For the specific examples shown here the largest stellar masses are up to $\sim 10^4 $ \msolar (e.g. in the \Rapid halo). However, the number densities are up to $\sim 10^5$ times higher than in the monolithic case making it possible to explain the number density of the whole massive BH population. 
 \item \textbf{Light Seeds:} Smaller haloes or those enriched with metals will produce relatively less massive stars. These stars will be distributed inhomogeneously around the halo centre and will accrete very inefficiently. The largest stellar masses shown here have masses up to a few times $\sim 10^2 $ \msolar (PopIII halo). The number densities of these halos are between  $10^{-1}$ cMpc$^{-3}$ and up to $10$ cMpc$^{-3}$ (z $\sim$ 10).
 
 \end{enumerate}

Based on these findings and results of other investigations, we speculate on  seed \textit{growth}  as follows:

\begin{enumerate}

\item \textbf{Need for gas replenishment:} In all three halos considered the accretion rate on the stellar progenitors of the seeds drops after $\sim 10^5$ years as the readily accretable gas in the parent halo is consumed. Gas replenishment is needed \citep{Johnson_2007, Lupi_2024} through either mergers with gas-rich halos or accretion from the cosmic web.

 \item \textbf{Light Seed Accretion:} \textit{Light} seeds accrete inefficiently because of a combination of a small Bondi radius \citep{Pacucci_2017} and their collective dynamics that leads them to move randomly in the inhomogenous stellar+dark matter distribution \citep{Smith_2018, Pfister_2019}. In order to overcome this barrier, some mechanism is necessary to allow highly efficient and sustained accretion to take place: for instance if they are embedded in a star cluster \citep{Alexander_2014, Natarajan_2020}, are captured/capture gas clumps \citep{Lupi_2016} or a later merger brings them into an extremely gas rich region \citep{Johnson_2007,Pezzulli_2016}. The question is for how long the BHs are able to remain within this dense region?
 
  \item \textbf{Heavy Seed Accretion:} 
  \textit{Heavy} seeds experience the same bottlenecks as  \light seeds, but their effect is mitigated in relation to the seed masses: Bondi accretion scales as $M_{BH}^2$ and gravity scales as $M_{BH}$. A 10-times more massive seed has better chances of remaining in dense regions and accreting from them \citep{Pfister_2019, Lupi_2024}. 
 
 \item \textbf{Light Seed Mergers:}  The more erratic dynamics and small cross-section of  \light seeds will hamper their growth by BH-BH mergers except when they are located in dense stellar environments \citep{Reinoso_2023, Arca-Sedda_2023, Rantala_2024}. After a galaxy merger existing seeds with masses below $\sim 10^5$ \msolar are highly unlikely to bind in binaries \citep{Pfister_2019,Ma_2021}. 
 
  \item \textbf{Heavy Seed Mergers:}  similarly to what was described for accretion, \heavy seeds are favored proportionally to their mass, since gravity-related process are eased as the BH mass increases, but even \heavy seeds have a hard time reaching the ``center'' of the shallow and granular potential of high-redshift galaxies \citep{Ma_2021}. 
   
 \end{enumerate}

\noindent In summary \heavy seeds production is more likely the more symmetric the halo is. The advance of more sophisticated models and simulations over the last two decades in particular tracks this idea well. Initially the relatively coarse simulations which also lacked advanced cooling algorithms showed the formation of central massive stars \citep[e.g.][]{Abel_2000, Bromm_1999}. The inclusion of more sophisticated simulations at increased resolution has in the vast majority of cases led to more fragmentation and a more complex morphology within collapsing haloes, for both \light and \heavy seed scenarios. 

While \heavy seed formation has still been shown to be viable in these more sophisticated simulations the most massive objects formed tend to be lighter than the maximum mass (theoretically) possible. Moreover, while \light seeds will be abundant with characteristic masses in the range of a few tens of solar masses they will be scattered around a morphologically complex halo. Rapid growth in such an environment will be extremely unlikely. Indeed the growth of \heavy seeds in such an environment is still challenging albeit mitigated by the larger Bondi radius which scales as $\rm{M_{BH}^2}$.

So then the question becomes - what seeds will grow and within what environment? Rapid growth will only become viable if a seed (light or heavy) can sink towards and remain in the galaxy centre where the gas densities are highest. The heavier the initial mass the more likely this is to occur. \textit{Light} seeds, in isolation, are not likely to do this, but there are regions of the parameter space that allow for their growth. \cite{Smith_2018} showed in their study that out of 20,000 \light seeds none grew: this suggests that the fraction of \light seeds able to grow is less than $10^{-4}$.  More optimistically \cite{Shi_2024} find that the fraction of light seeds that grow, under favourable and somewhat idealised assumptions, is $\sim 10^{-2}$.  If the \light seed that grows is able to reach the typical mass of a \heavy seed, and the abundance of \light seeds is at least $10^{3}-10^{4}$ times that of \heavy seeds then \light seeds become equally likely to be the source of MBHs.  A point of note here is that it is currently impossible to constrain seed models based on current observational data of MBH masses by extrapolating the growth backwards. The available parameter space, which allows both Eddington and super-Eddington phases makes this an impossible challenge. All that can be garnered from current observational data is an upper limit on the seed masses.\\

\begin{figure*}
\centerline{
  \includegraphics[width=17.0cm, height=13.0cm]{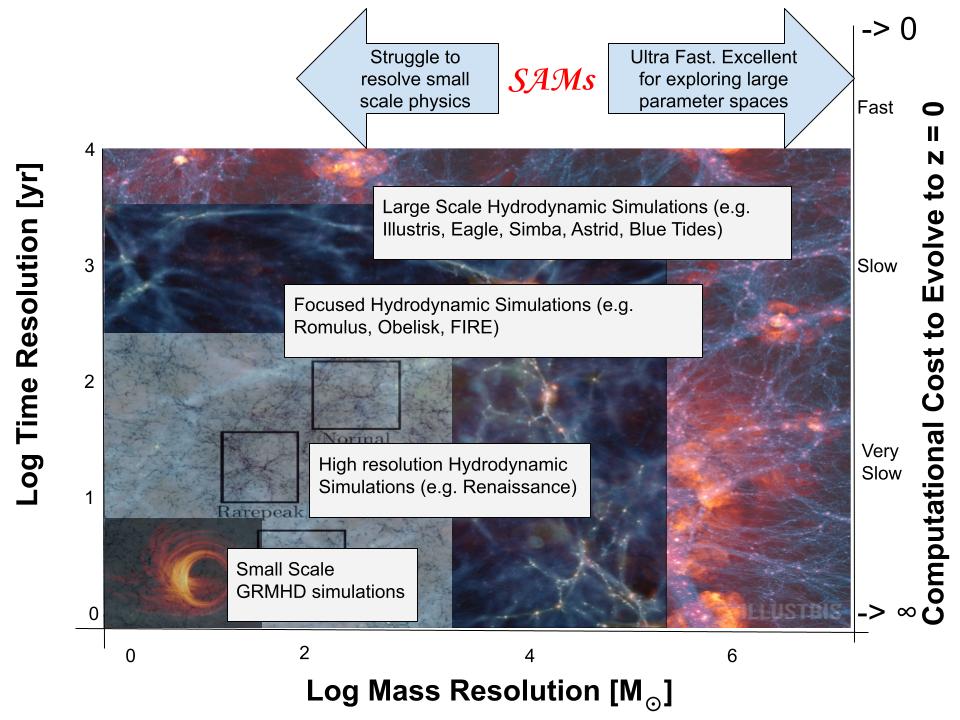}}
\caption{Schematic comparing Semi Analytic Models (SAMs) against hydrodynamic simulations. The computational complexity increases from the top right to the bottom left. Models in the top right run extremely quickly but can lack the physical complexity to resolve small scale features like MBH dynamics and seed formation. On the other hand models occupying the bottom left panel focus on 
scales close to the event horizon and can only be evolved for relatively short physical times given the extreme computational expenses. An array of simulations volumes and resolutions must therefore be used to "join" the scales. The simulation images shown include Illustris \citep[e.g.][]{Vogelsberger_2014}, Romulus \citep[e.g.][]{Tremmel_2017}, Renaissance \cite[e.g.][]{Xu_2013} and results from small scale GRMHD simulations \citep{Chatterjee_2019}.}
\label{Fig:ModelSchematic}
\end{figure*}

\section{Outlook}
\noindent The most recent generation of large scale hydrodynamical simulations are beginning to achieve mass and spatial resolutions at a level where self-consistently tracking light and heavy seed formation and growth becomes tractable. As discussed in \S \ref{Sec:Growth} the Astrid, NewHorizon and Romulus simulation suites already model \heavy seed-like objects and study their growth over cosmic timescales. While the 
formation of the heavy seeds from either a stellar or collisionally driven pathway is not followed self-consistently the subsequent growth 
through accretion and mergers of the MBH is followed reasonably accurately. Moreover, the simulations include subgrid models for dynamical friction which model the dynamics of these \heavy seed black holes within growing proto-galaxies.  \\
\indent Semi Analytic Models (SAMs) rely on a backbone of dark matter merger trees  to model galaxy and MBH evolution over long timescales. Their relative simplicity, compared to hydrodynamic simulations make them ideal for rapidly traversing the large parameter space of galaxy and black hole formation \citep[e.g.][]{Schneider_2006, Dayal_2020, Trinca_2022}.  However, this power comes at the cost of small scale details. SAMs inherently do not have spatially-resolved structures and the detailed physical modelling capacity necessary to capture emergent physics like turbulence, detailed ISM physics as well as the likely episodic nature of MBH accretion. \\
\indent In Figure \ref{Fig:ModelSchematic} we show a schematic outlining the range of spatial and temporal scales that must be bridged by our models. No one simulation suite, technique or code can cross the domains and multiple suites and techniques (SAMs + hydrodynamics) must be used. To model MBH evolution different models 
utilise different seeding and accretion prescriptions - usually driven by the 
numerical resources available to them at a given scale. For example the large scale simulations (top right) must seed galaxies with relatively large black holes which represent black holes not necessarily at their seed masses but at some time after their physical formation. \\
\indent SAMs based on high-resolution parent DM halo trees can in principle seed MBHs in arbitrarily low-mass galaxies/halos. However, SAMs have to seed BHs based on global quantities rather than local, spatially-resolved properties. Common to both the large scale hydrodynamic simulations and the SAMs is a difficulty in resolving the dynamics and detailed accretion physics of the MBHs. As we negotiate the parameter space of the schematic moving from top right to bottom left the models can becomes increasingly sophisticated in how, and at what mass, black holes are seeded. At smaller scales and higher resolution emergent physics gives rise to a better handle on the physics thought necessary for MBH formation (be they light or heavy seeds). Conversely, these simulations are unable, to follow over long timescale, the evolution of the seeds.
\\
\indent The next generation of simulations are likely to push this mass resolution limit even further slowing encroaching into the regime of the \light seed (simulations by \cite{Bhowmick_2024, Bhowmick_2024a} already incorporate some of this thinking, albeit with optimistic dynamics as BHs are pinned to halo centres, increasing the chances of growth by both accretion and mergers). By including a self-consistent treatment of both \light and \heavy seeds and their  dynamics, the growth and relative number densities of each seed population can in principle be determined. 
However, expectations must be tempered. For the foreseeable future a subgrid model for both \light and \heavy seed formation will exist particularly as the formation pathways for \heavy seeds via collisional runaway \citep[e.g.][]{Arca-Sedda_2023, Rantala_2024} and via SMSs remain unconverged in detailed high resolution simulations \citep[e.g.][]{Prole_2024}.\\
\indent From the observational point of view, JWST has detected a new population of AGN, selected spectroscopically via broad \citep{Maiolino_2023, Matthee_2024} or narrow emission lines \citep{Scholtz_2023}, or photometrically via color-color selection and then followed up spectroscopically to identify broad emission lines \citep{Greene_2024}. This population has a number density above the extrapolation of the previously-known bright end of the AGN luminosity function: this is because these AGN have little or no X-ray emission \citep{Furtak_2024,2024arXiv240413290Y} and they are  UV-faint with respect to the host galaxies \citep{Matthee_2024,Scholtz_2023}. The physical reason for this is still unclear, but previous selections were based on X-ray and restframe-UV, and therefore this population was missed. Interestingly, reanalysis of HST images using variability has uncovered a similarly sized population that is UV bright \citep{Hayes_2024} and some of the high-z AGN are X-ray bright \citep{Bogdan_2023}. However, as a population, the JWST AGN, or perhaps candidates, do not show X-ray emission \citep{Yue_2024,Maiolino_2024Xray,Ananna_2024}. This lack of X-rays has led to different hypotheses: the sources are not AGN, they're AGN but the masses of MBHs have been overestimated, or they are intrinsically X-ray weak, or they are surrounded by dust-free Compton-thick gas. 

These sometimes conflicting results make it difficult to anchor theoretical models to observations, and in these early days of these new JWST results we may consider taking an agnostic attitude and perhaps consider the AGN as candidates, while waiting for additional in-depth analyses. No doubt, however, this influx of data to previously inaccessible wavelengths has opened new pathways to test and constrain theoretical models of MBH formation and growth by accretion. In the not too far future, LISA and ET will probe MBH growth by mergers for massive and intermediate-mass BHs up to very high redshift \citep{eLISA,Maggiore_2020}.\\

\section*{Acknowledgements}
\noindent We thank the referee for providing highly constructive feedback which greatly improved the manuscript.  JR acknowledges support from the Royal Society and Science Foundation Ireland under grant number 
 URF\textbackslash R1\textbackslash 191132. JR also acknowledges support from the Irish Research Council Laureate programme under grant number IRCLA/2022/1165. MV  acknowledges funding from the French National Research Agency (grant ANR-21-CE31-0026, project MBH\_waves).

\label{lastpage}
\bibliographystyle{mn2e}
\bibliography{mybib}
\end{document}